\documentclass[a4paper]{jpconf}
\usepackage{graphicx}
\usepackage[numbers,square,sort&compress]{natbib}
\begin{document}
\title{Crystal structure and thermodynamic properties of the non-centrosymmetric PrRu$_4$Sn$_6$ caged compound}

\author{Michael O Ogunbunmi and Andr\'{e} M Strydom}

\address{Highly Correlated Matter Research Group, Physics Department, University of Johannesburg, P. O. Box 524,  Auckland Park 2006, South Africa.}

\ead{moogunbunmi@gmail.com, mogunbunmi@uj.ac.za}

\begin{abstract}
 PrRu$_4$Sn$_6$ is a tetragonal, non-centrosymmetric structure compound. It is isostructural to the extensively studied Kondo insulator CeRu$_4$Sn$_6$ which crystallizes in the YRu$_4$Sn$_6$-type structure with space group $I$\={4}2$m$. In this structure, the Pr atom fills the void formed by the octahedral Ru$_4$Sn$_6$ units which results in a tetragonal body-centred arrangement. Here we present reports on the physical and magnetic properties of PrRu$_4$Sn$_6$. The temperature dependences of specific heat, $C_p(T)$, electrical resistivity, $\rho(T)$, and magnetic susceptibility, $\chi(T)$, reveal the absence of a long-range magnetic ordering down to 2 K. $\chi(T)$ follows a Curie-Weiss behaviour above 100~K with an effective magnetic moment, $\mu_\mathrm{eff}$ = 3.34~$\mu_B$/Pr and paramagnetic Weiss temperature, $\theta_p$ = $-$19.47~K indicating a dominant antiferromagnetic interaction. The magnetization at 2~K is quasi-linear in nature and attains a value of 0.86~$\mu_B$/Pr at 7 T which is well reduced compared to the calculated value of 3.32~$\mu_B$/Pr expected for a free Pr$^{3+}$ ion. This is attributed to possible magneto-crystalline anisotropy in the system. $C_p(T)$ indicates the presence of a optical-phonon mode which is supported by a glass-like thermal conductivity above $\sim$45~K. This observation is associated with caged structured compounds where the low-frequency optical-phonon mode of the guest atom interacts with the host lattice, resulting in the scattering of heat-carrying quasiparticles. 
\end{abstract}

\section{Introduction}
The $R$Ru$_4$Sn$_6$ ($R$ = Y, La-Nd, Sm, Gd-Ho) series are intermetallic compounds which crystallize in the tetragonal YRu$_4$Sn$_6$-type structure with  a non-centrosymmetric space group $I$\={4}2$m$ (No. 121) \cite{zumdick1999condensed}. The structure was first reported by Venturini \emph{et al} \cite{venturini1990crystal}. The crystal structure is made up of an octahedral Ru$_4$Sn$_6$ unit enclosing the guest $R$ atom. Crystal structures of this nature have generated much interest lately especially in the search for new superconductors \cite{yamada2018superconductivity,winiarski2016rattling}. Also, the non-centrosymmetric nature of the space group is characteristic of certain superconductors where the mixing of the spin-singlet and spin-triplet Cooper pairing channels have been found to give rise to a two-component order parameter \cite{bauer2004heavy,bauer2009baptsi,okuda2007magnetic}. CeRu$_4$Sn$_6$ is a Kondo insulator, and it is the most extensively studied member of the series \cite{strydom2005electronic,bruening2006119sn,bruning2010low,sundermann2015ceru}. Other studies by Koch and Strydom reveal a magnetic ordering for the isostructural compounds of $R$Ru$_4$Sn$_6$, with $R$ = Sm, Gd and Dy compounds at low temperatures while those of Nd, Tb and Ho compounds are paramagnetic down to 2~K \cite{koch2008electronic}. \\
\indent
As part of our search for Pr-based systems exhibiting novel ground states, we have synthesized a polycrystalline sample of  PrRu$_4$Sn$_6$ and investigated its physical and magnetic properties. It is noted that the existence of PrRu$_4$Sn$_6$  was first reported by Zumdick and P\"{o}ttgen \cite{zumdick1999condensed} but no physical or magnetic properties have been reported thereafter. The Pr atom in this structure has a tetragonal site symmetry of $D_{2d}$ similar to those of the Pr$_3T_4X_{13}$ compounds, resulting in the crystal electric field splitting of the $J$ = 4 multiplet into seven levels consisting of five singlets and two non-Kramers doublets. 

\section{Experimental methods}
\begin{figure}[!t]
	\centering
	
	\includegraphics[scale=1.1]{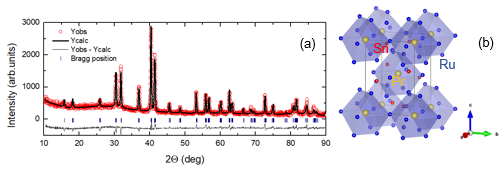}
	\caption{\label{fig_PrRu4Sn6_xrd} (a): Powder X-ray diffraction pattern of PrRu$_4$Sn$_6$ (red circles) with a Rietveld refinement (black line) based on the $I$\={4}2$m$ space group (No. 121). The vertical bars are the Bragg peak positions while the grey line represents the difference between the experimental and calculated intensities. (b): Crystal structure of PrRu$_4$Sn$_6$ showing Pr atom being enclosed by the Ru$_4$Sn$_6$ octahedral unit. }
\end{figure}

A polycrystalline sample of PrRu$_4$Sn$_6$ was prepared by arc melting stoichiometric amounts of high-purity elements (wt.\% $\ge$ 99.9) on a water-cooled Cu plate under a purified static argon atmosphere in an Edmund Buehler arc furnace. The weight loss after melting was $\sim$ 0.05\%. The arc-melted pellet was wrapped in Ta foil, placed in an evacuated quartz tube and annealed at 900$^\circ$C for 21 days. 
A powder X-ray diffraction (XRD) pattern was recorded on a pulverized sample using a Rigaku diffractometer employing Cu-K$\alpha$ radiation. The obtained powder XRD pattern was refined using the Rietveld method \cite{thompson1987rietveld} employing the FullProf suite of programs \cite{rodriguez1990fullprof}. We found that the compound was phase-pure within the limits of the resolution of the instrument. In Table~\ref{PrRu4Sn6_parameters}, the atomic positions and lattice parameters obtained from the refinement are presented and are comparable with a  previous report \cite{zumdick1999condensed}.  The refined XRD pattern and the crystal structure are shown in Fig.~\ref{fig_PrRu4Sn6_xrd}. \\
\indent
Magnetic properties were measured using the Magnetic Property Measurement System (Quantum Design Inc., San Diego) between 2~K and 300~K with an external magnetic field up to 7~T. The four-probe DC electrical resistivity, specific heat and thermal transport measurements between 2~K and 300~K were measured using the Physical Property Measurement System also from Quantum Design.
\begin{table}[h!]
		\centering

			\caption{The atomic positions and lattice parameters of PrRu$_4$Sn$_6$ obtained from a Rietveld refinement of the XRD pattern.}
			\label{PrRu4Sn6_parameters}
			\begin{tabular}{|c|c|c|c|c|c|c|}
				\hline
				Site notation &	Atom & Wyckoff site & Point symmetry & $x$ & $y$ & $z$ \\ \hline
				Sn(1) & Sn &  8$i$ & $m$ & 0.17635 & 0.17635 & 0.28771 \\
				Ru & Ru & 8$i$ & $m$ & 0.32788 & 0.32788 & 0.08126  \\
				Sn(2) & Sn & 4$c$ & 222 & 0 & 1/2 & 0  \\
				Pr & Pr &  2$a$ & $-$42$m$ & 0 & 0 & 0  \\
				
				\hline \hline
					$a$ (\AA) & $c$ (\AA) &  $V$ (\AA$^3$) & formula units ($Z$)  &$R_{wp}$ (\%) & $R_p$ (\%) & $\chi^2$  \\  \hline
					6.870(3) & 9.761(2) & 461.5(9) & 2 & 8.588 & 7.295 & 5.210 \\ \hline

			\end{tabular}
	
	\end{table}
\section{Magnetic properties} 

\begin{figure}[!t]
	\centering
	\includegraphics[scale=0.55]{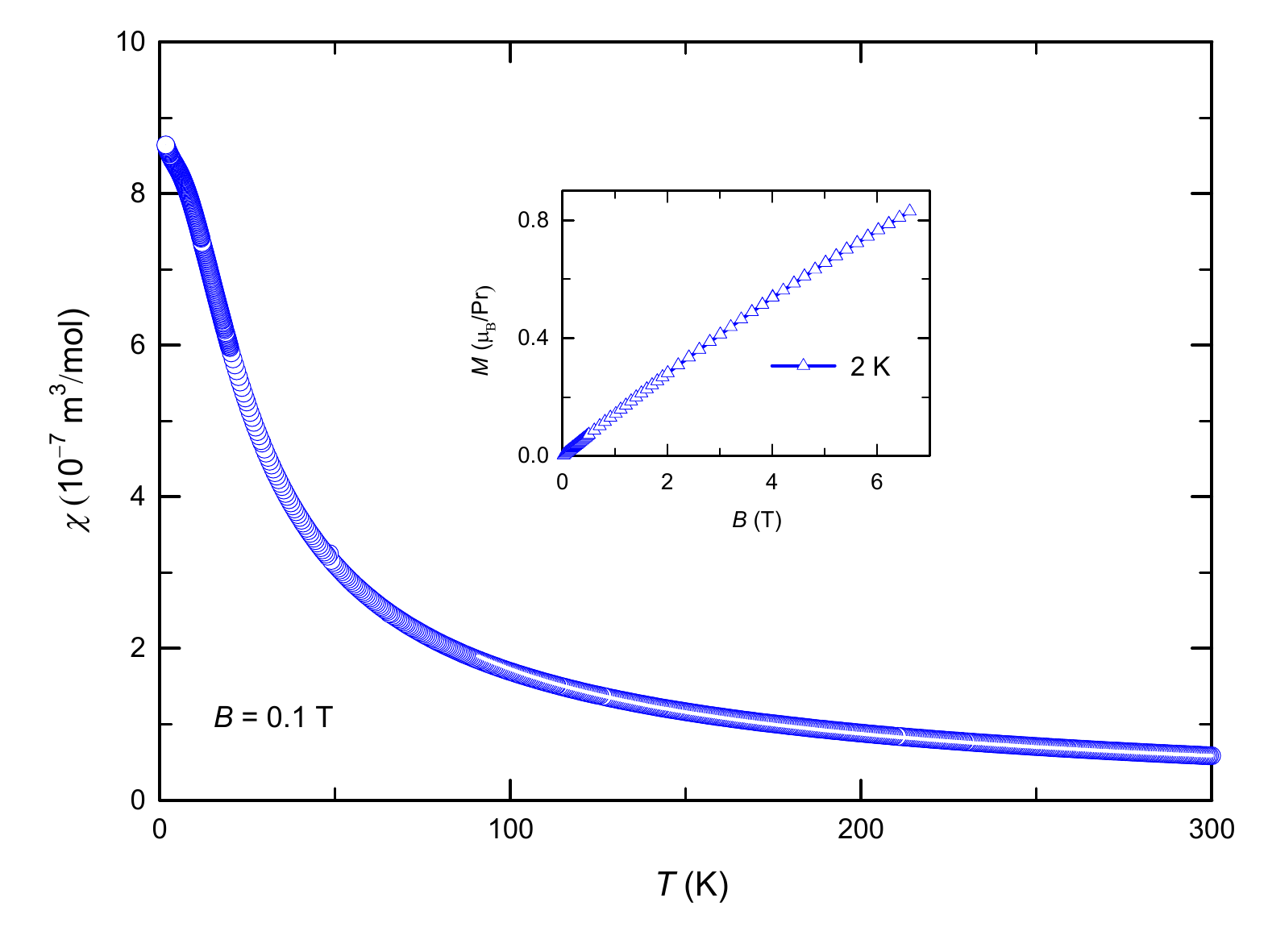}
	\caption{\label{fig_PrRu4Sn6_chi} Temperature dependence of magnetic susceptibility, $\chi(T)$, of PrRu$_4$Sn$_6$ measured in a field of 0.1~T. The white-solid line is a Curie-Weiss fit described in the text. Inset (b): Isothermal magnetization of PrRu$_4$Sn$_6$ at 2~K.}
\end{figure}

The temperature dependence of magnetic susceptibility, $\chi(T)$, of PrRu$_4$Sn$_6$ in an external field of 0.1~T and in the temperature range of 2~K to 300~K is presented in Fig.~\ref{fig_PrRu4Sn6_chi}. $\chi(T)$ shows a paramagnetic behaviour down to low temperatures with no indication of a long-range magnetic ordering observed. The white-solid line is a Curie-Weiss fit based on the expression: $\chi(T) = N_A \mu_\mathrm{eff}^2/(3k_B(T-\theta_p))$ for data above 100~K with values of effective magnetic moment, $\mu_\mathrm{eff} = 3.34~\mu_\mathrm{B}$/Pr and Weiss temperature, $\theta_p = -19.47~$K. The observed $\mu_\mathrm{eff}$ is close to the calculated value of 3.58~$\mu_\mathrm{B}$/Pr expected for a free Pr$^{3+}$ ion. At low temperatures, a Van-Vleck paramagnetic behaviour in $\chi(T)$ suggests a nonmagnetic ground state in PrRu$_4$Sn$_6$. The isothermal magnetization at 2~K is presented in the inset (b) of Fig.~\ref{fig_PrRu4Sn6_chi}. The magnetization follows a quasi-linear behaviour up to 7~T and attains a value of 0.86~$\mu_\mathrm{B}$/Pr at 7~T which is well reduced compared to the saturation moment of 3.32~$\mu_\mathrm{B}$/Pr expected for a free Pr$^{3+}$ ion implying a possible magneto-crystalline anisotropy in the compound. 

\section{Specific heat}
\begin{figure}[t!]
	\centering
	\includegraphics[scale=0.55]{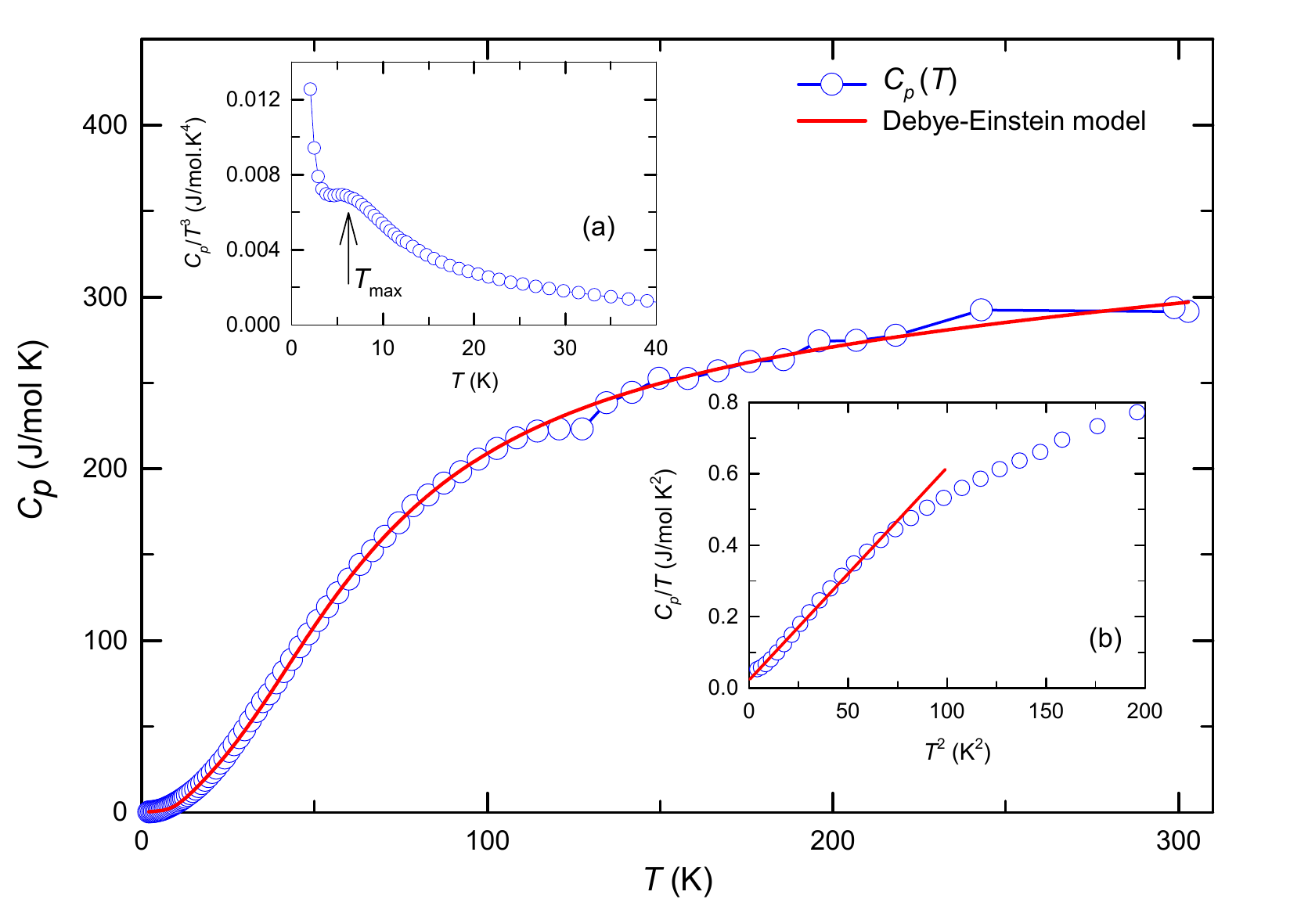}
	\caption{\label{fig_PrRu4Sn6_cp} Temperature dependence of specific heat, $C_p(T)$, of PrRu$_4$Sn$_6$. Inset (a): Low-temperature plot of $C_p/T^3$ against $T$. Inset (b): Plot of $C_p/T$ against $T^2$ along with a linear fit indicated by the red-solid line to extract the Sommerfeld coefficient.}
\end{figure}

The temperature dependence of specific heat, $C_p(T)$, of PrRu$_4$Sn$_6$ studied between 2~K and 300~K is presented in Fig.~\ref{fig_PrRu4Sn6_cp}. Inset (a) of Fig.~\ref{fig_PrRu4Sn6_cp} shows a plot of $C_p/T^3$ against $T$. Such a plot is important in determining the possible presence of low-frequency Einstein modes in $C_p(T)$ through the occurrence of a local maximum in $C_p/T^3$. A local minimum is observed in the plot as indicated by the arrow at $T_{max}$ = 6~K which confirms the presence of low-frequency Einstein modes in PrRu$_4$Sn$_6$. $T_\mathrm{max}$ is the temperature below which the Einstein modes are frozen out. By using a model incorporating both the Debye and Einstein terms, the experimental specific heat is fitted as shown by the red line in Fig.~\ref{fig_PrRu4Sn6_cp}. The Debye-Einstein model is given by:

\begin{equation}
C_p(T) = mD\left(\frac{\theta_D}{T}\right) + nE\left(\frac{\theta_E}{T}\right),
\end{equation}

\begin{equation}
D\left(\frac{\theta_D}{T}\right) = 9R\left(\frac{T}{\theta_D}\right)^3 \int_{0}^{\theta_D/T}  \frac{ x^4\exp(x)}{(\exp(x) - 1)^2}dx,
\end{equation}

\begin{equation}
E\left(\frac{\theta_E}{T}\right) = 3R\left(\frac{\theta_E}{T}\right)^2\cdot\frac{\exp\left(\theta_E/T\right)}{(\exp\left(\theta_E/T\right) - 1)^2},
\end{equation}
where $\theta_D$ and $\theta_E$ are the Debye and Einstein temperatures with values of 241.73(9)~K and 32.431(3)~K, respectively. It is observed that $T_{max}$ $\simeq$ 0.2$\theta_E$ which is in agreement with the observation in Ce$_3$Rh$_4$Sn$_{13}$ \cite{kohler2007low}. In Inset (b), a plot of $C_p/T$ against $T^2$ is shown together with a least-square fit (red line) based on the expression: $C_p/T = \gamma + \beta T^2$ and $\beta = 12\pi^4nR/(5\theta^3_D$), where $n$ and R are the number of atoms per formula unit and universal gas constant, respectively, $\gamma$ is the Sommerfeld coefficient and $\theta_D$ is the Debye temperature. Values obtained from the fit are: $\gamma = 38.60~$mJ/(K$^2$ mol) and $\theta_D$ = 154.50~K. The $\gamma$ observed for PrRu$_4$Sn$_6$ is about 10 times the values found in ordinary metals. 

\section{Transport properties}
\begin{figure}[!t]
	\centering
	\includegraphics[scale=0.75]{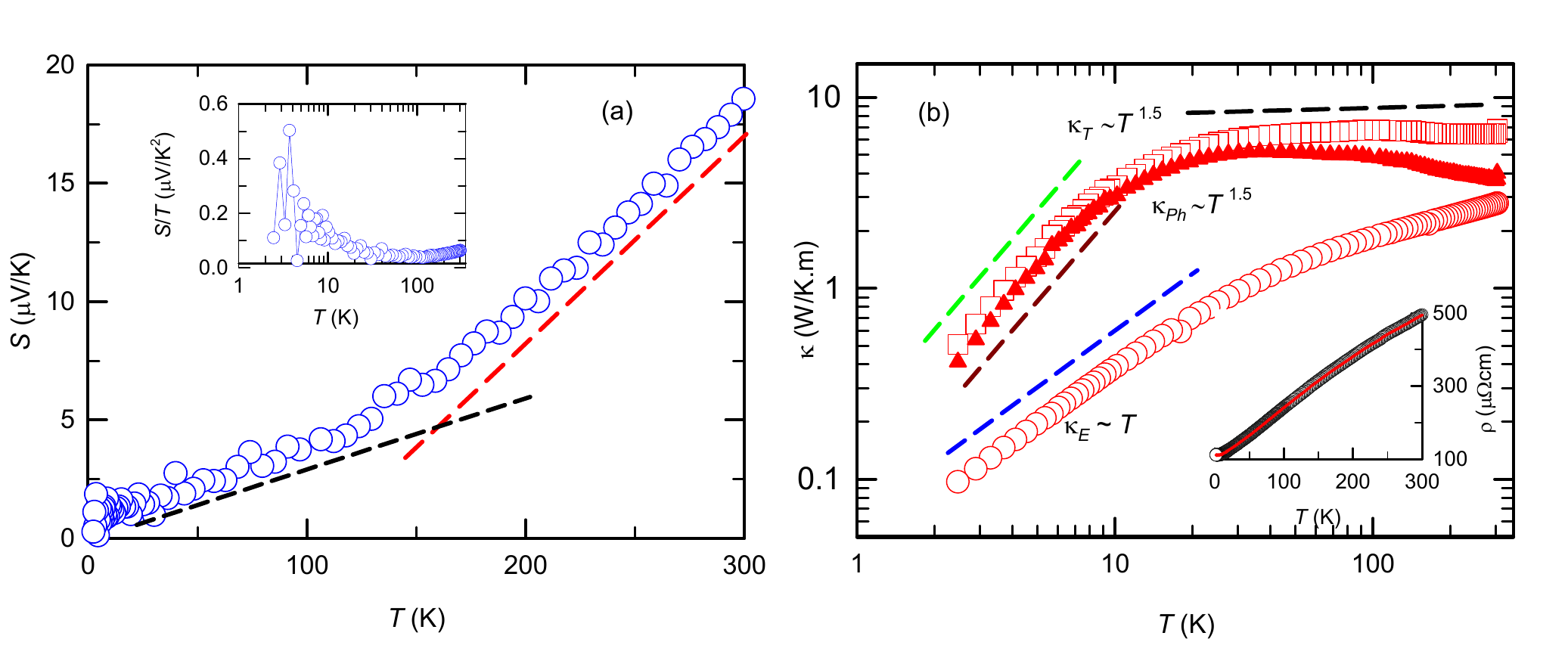}
	\caption{\label{fig_PrRu4Sn6_tto} (a) Temperature dependence of thermoelectric power, $S(T)$, of PrRu$_4$Sn$_6$. The red and black-dashed lines are guides to the eye, indicating a change in slope of $S(T)$ at $\sim$145~K. Inset: Plot of $S/T$ against $T$ on a semi-log axis. (b) Temperature dependences of total thermal conductivity, $\kappa_T(T)$, phonon thermal conductivity, $\kappa_{Ph}(T)$  and electronic thermal conductivity, $\kappa_E(T)$. The green, brown and blue dashed-lines represent the power-law behaviours of $\kappa_T(T)$,  $\kappa_{Ph}(T)$ and $\kappa_E(T)$, respectively while the black-dashed line is a guide to the eye described in the text. Inset:  Temperature dependence of electrical resistivity with a BG fit (red line) described in the text.}
\end{figure}

To further understand the physical properties of PrRu$_4$Sn$_6$, a thermal transport measurement was carried out between 2~K and 300~K. The temperature dependences of thermoelectric power, $S(T)$, and thermal conductivity, $\kappa(T)$, were measured simultaneously on a bar-shaped sample. As shown in Fig.~\ref{fig_PrRu4Sn6_tto} (a), $S(T)$ is positive throughout the temperature range investigated and attains a value of 18.81~$\mu$V/K at room temperature. The red and black-dashed lines suggest two areas of linear-in-$T$ behaviour on either side of $\sim$135~K. At 2~K, $S(T)$ has a value of $\sim$1~$\mu$V/K indicating a significant drop in the carrier concentration between room temperature and 2~K. The change in slope of $S(T)$ at about 145~K is consistent with the anomaly observed in $C_p(T)$ around the same temperature. The origin of such an observation is not immediately clear and further measurements are needed to resolve the physics at play. A plot of $S(T)/T$ is shown in the inset of Fig.~\ref{fig_PrRu4Sn6_tto} (a). For $T \le 100$~K the slope of $S(T)/T$ is ~$\sim$ 0.7~$\mu$ V/K$^2$ which is slightly above those of ordinary metals. The general feature of $S(T)$ suggests a hole-type charge carriers near the Fermi level. \\
\indent
The total thermal conductivity, $\kappa_T(T)$, of PrRu$_4$Sn$_6$ is presented in Fig.~\ref{fig_PrRu4Sn6_tto} (b) on a $\log$-$\log$ axes. $\kappa_T(T)$ is nearly temperature independent from room temperature down to about 45~K (as shown by the black-dashed line) which is characteristic of a glassy behaviour in thermal conductivity. The observation of a glass-like thermal conductivity in a crystalline compound is often associated with caged systems. The low-frequency optical-phonon mode of the guest atom scatters heat-carrying quasiparticles thus leading to a reduction in lattice thermal conductivity. Using the Wiedemann-Franz relation \cite{kittel1996introduction} given as: $\kappa = L_0 T/\rho(T)$, where $L_0$ is the Lorentz number given by: $L_0$ = $\pi^2k_ B^2/3e^2$ = $2.45 \times 10^{-8}~$W$\Omega$/K$^{2}$, the electronic contribution to the thermal conductivity, $\kappa_E(T)$ is extracted and it is also presented in Fig.~\ref{fig_PrRu4Sn6_tto} (b). Also shown in the plot is $\kappa_{Ph}(T)$ obtained by subtracting $\kappa_E(T)$ from $\kappa_T(T)$.  Below about 10~K, $\kappa_T(T)$ and $\kappa_{Ph}(T)$ show power-law behaviour of $T^{1.5}$ while  $\kappa_E(T)$ is linear-in-$T$ as indicated by the green, brown  and blue-dashed lines. This indicates a good metallic behaviour.  $\kappa_{Ph}(T)$ $>$ $\kappa_E(T)$  in the whole temperature range studied revealed that the heat transport is not charge-carrier dominated.  \\
\indent
The temperature dependence of electrical resistivity, $\rho(T)$, of PrRu$_4$Sn$_6$ is presented in the inset of Fig.~\ref{fig_PrRu4Sn6_tto} (b) between 2~K and 300~K. $\rho(T)$ follows a typical metallic behaviour down to low temperature with residual resistivity ratio $\approx$ 5 which indicates a good crystalline quality. No signature of long-range magnetic or any type of ordering is observed in the temperature range studied in support of the observations in $\chi(T)$ and $C_p(T)$. To further understand the electrical transport properties of PrRu$_4$Sn$_6$, the Bloch-Gr\"{u}neisen (BG) expression \cite{mott1958theory} was fitted to the data in the whole temperature range (shown as a red line). The BG expression is given as:

\begin{equation}
\rho(T) = \rho_0 + \frac{4K}{\Theta_R}\left(\frac{T}{\Theta_R}\right)^5\int_{0}^{\Theta_R/T} \frac{x^5 dx}{(e^x -1)(1-e^{-x})}, 
\end{equation}
	
where $\rho_0$ is the residual resistivity due to defect scattering in the crystal lattice, $K$ is the electron-phonon coupling constant and $\Theta_R$ is the resistivity Debye temperature. Values of $\rho_0$ =  102.8(2)~$\mu \Omega$ cm, $K$ =  90.19(1)~$\mu \Omega$ cm K, and $\Theta_R = 39.20(1)$~K  are obtained from the least-square fit.This observation here further supports a metallic behaviour of PrRu$_4$Sn$_6$.

\section{Conclusion}
We have studied the physical and magnetic properties of the non-centrosymmetric PrRu$_4$Sn$_6$ compound. A paramagnetic ground state is inferred from the magnetic susceptibility results down to 2~K. The presence of low-frequency Einstein modes are observed in $C_p(T)$. This observation is further supported by the glass-like thermal conductivity for temperatures above 45~K. $S(T)$ undergoes a change in slope at $\sim$ 145~K, which is around the same temperature an anomaly in $C_p(T)$ is observed.  Further measurements are expected to help clarify the origin of the observations in $C_p(T)$ and $S(T)$.\\

\section*{Acknowledgement}
MOO acknowledges the UJ-URC bursary for doctoral studies in the Faculty of Science. AMS thanks the SA-NRF (93549) and UJ-URC for financial support.


\bibliographystyle{iopart-num}
\bibliography{PrRu4Sn6}

\end{document}